\documentclass{sig-alternate}

\usepackage{url}

\newcommand{\E}{{\mathsf E}}
\newcommand{\Atil}{{\widetilde{A}}}
\renewcommand{\SS}{{\mathcal{S}}}
\newcommand{\BB}{{\mathcal{B}}}

\newcommand{\ZZ}{{\mathbb{Z}}}

\newcommand{\DD}{{\mathcal D}}

\newcommand{\KK}{{\mathcal K}}
\newcommand{\HH}{{\mathcal H}}
  
\newcommand{\F}{{\sf F}} 
\newcommand{\diag}{{\rm diag}}
\newcommand{\so}{{O {\;\!\tilde{}}\,}}

\newcommand{\nxn}{{n\times n}}
\newcommand{\nx}[1]{{n\times #1}}
\newcommand{\norm}[1]{\|#1\|}
\newcommand{\divs}{{\mskip3mu|\mskip3mu}}

\newcommand{\litem}{\item[]\noindent\kern-\leftmargin}
\newcommand{\rddots}{\mathinner{
    \mkern2mu\raise1pt\hbox{.}
    \mkern2mu\raise4pt\hbox{.}
    \mkern1mu\raise7pt\vbox{\kern7pt\hbox{.}}
    \mkern1mu}}
\newcommand{\softO}{{O\mskip1mu\tilde{\,}\mskip1mu}}
\newcommand{\KKul}{{\KK_u^{(\ell)}}}
\newcommand{\KKur}{{\KK_u^{(r)}}}
\newcommand{\Kul}{{K_u^{(\ell)}}}
\newcommand{\Kur}{{K_u^{(r)}}}

\newtheorem{theorem}{Theorem}[section]

\newtheorem{corollary}[theorem]{Corollary}

\newtheorem{proposition}[theorem]{Proposition}

\newtheorem{remark}[theorem]{Remark}

\newcommand{\rank}{\textup{rank}\,}
\numberwithin{equation}{section}
\numberwithin{table}{section}

\newcount\mn \newcount\hr
\def\timeofday{
  \hr=\time \divide\hr 60
  \mn=-\hr \multiply\mn 60 \advance\mn \time
  \ifnum\hr=0
     {12\,:\,\twodigits\mn\,am}
  \else{
     \ifnum\hr<13
        {\number\hr\,:\,\twodigits\mn\,am}
     \else
        {\advance\hr -12 \number\hr\,:\,\twodigits\mn\,pm}
     \fi}
  \fi}
\def\twodigits#1{\ifnum #1<10 0\fi \number#1}
\begin{document}

\title{Faster Inversion and Other Black Box Matrix Computations
Using Efficient Block Projections} %

\numberofauthors{1}

\newfont{\adresse}{phvr at 9pt}
\author{
%
  \alignauthor  
  Wayne Eberly$^1$, 
  Mark Giesbrecht$^2$, 
  Pascal Giorgi$^{2,4}$, 
  Arne Storjohann$^2$, 
  Gilles Villard$^3$\\
  \affaddr{~~~}\\
  \adresse{(1) Department of Computer Science, U. Calgary}\\
  \adresse{\url{http://pages.cpsc.ucalgary.ca/~eberly}}\\[3pt]
  \adresse{(2) David R. Cheriton School of Computer Science, U. Waterloo} \\
  \adresse{\url{http://www.uwaterloo.ca/~{mwg,pgiorgi,astorjoh}}}\\[3pt]
  \adresse{(3) CNRS, LIP, \'Ecole Normale Sup\'erieure de Lyon}\\
  \adresse{\url{http://perso.ens-lyon.fr/gilles.villard}}\\[3pt]
  \adresse{(4) IUT - Universit\'e de Perpignan}\\
  \adresse{\url{http://webdali.univ-perp.fr/~pgiorgi}}
}
\maketitle

\renewcommand{\thefootnote}{\fnsymbol{footnote}}

\footnotetext[1]{This material is based on work supported in part by
  the French National Research Agency (ANR Gecko, Villard), and by the
  Natural Sciences and Engineering Research Council (NSERC) of Canada
  (Eberly, Giesbrecht, Storjohann). 
}

\begin{abstract}
  Efficient block projections of non-singular matrices have recently
  been used, in \cite{EGGSV06-1}, to obtain an efficient algorithm to
  find rational solutions for sparse systems of linear equations. In
  particular a bound of $\softO(n^{2.5})$ machine operations is
  obtained for this computation assuming that the input matrix can be
  multiplied by a vector with constant-sized entries using $\softO(n)$
  machine operations. Somewhat more general bounds for black-box
  matrix computations are also derived. Unfortunately, the correctness
  of this algorithm depends on the existence of efficient block
  projections of non-singular matrices and this has been conjectured
  but not proved.

  In this paper we establish the correctness of the algorithm
  from~\cite{EGGSV06-1} by proving the existence of efficient block
  projections for arbitrary non-singular matrices over sufficiently
  large fields. We further demonstrate the usefulness of these
  projections by incorporating them into existing black-box matrix
  algorithms to derive improved bounds for the cost of several matrix
  problems --- considering, in particular, ``sparse'' matrices that
  can be be multiplied by a vector using $\softO(n)$ field operations:
  We show how to compute the dense inverse of a sparse matrix over any
  field using an expected number of $\softO(n^{2.27})$ operations in
  that field.  A basis for the null space of a sparse matrix --- and a
  certification of its rank --- are obtained at the same cost. An
  application of this technique to Kaltofen and Villard's
  Baby-Steps/Giant-Steps algorithms for the determinant and Smith Form
  of an integer matrix is also sketched, yielding algorithms requiring
  $\so(n^{2.66})$ machine operations. More general bounds involving
  the number of black-box matrix operations to be used are also
  obtained.

  The derived algorithms are all probabilistic of the Las Vegas
  type. That is, they are assumed to be able to generate random
  elements from the field at unit cost, and always output the correct
  answer in the expected time given. 
\end{abstract}

\section{Introduction}

In our paper \cite{EGGSV06-1} we presented an algorithm which
purportedly solved a sparse system of rational equations considerably
more efficiently than standard linear equations solving.
Unfortunately, its effectiveness in all cases was conjectural, even as
its complexity and actual performance were very appealing.  This
effectiveness relied on a conjecture regarding the existence of
so-called efficient block projections.  Given a matrix $A\in\F^\nxn$ over
any field $\F$, these projections should be block vectors
$u\in\F^{n\times s}$ (where $s$ is a blocking factor dividing $n$, so
$n=ms$) such that we can compute $uv$ or $v^tu$ quickly for any
$v\in\F^{n\times s}$, and such that the sequence of vectors $u$, $Au$,
\ldots, $A^{m-1}u$ has rank $n$.

In this paper, we prove the existence of a class of such efficient block
projections for non-singular $n \times n$ matrices over sufficiently
large fields --- we require that the size of the field~$\F$ exceed
$n(n+1)$.  This implies our algorithm from \cite{EGGSV06-1} for
finding the solution to $A^{-1}b$ for a ``sparse'' system of equations
$A\in\ZZ^\nxn$ and $b\in\ZZ^\nx1$ works as stated using these
projections, and requires fewer bit operations than any previously
known when $A$ is sparse, at least using ``standard'' (i.e., cubic)
matrix arithmetic.  Here, by $A$ being sparse, we mean that it has a
fast matrix-vector product modulo any small (machine-word size) prime
$p$.  In particular, our algorithm requires an expected
$\so(n^{1.5}(\log(\norm{A}+\norm{b})))$ matrix-vector products
$v\mapsto Av\bmod p$ for $v\in\ZZ_p^\nx1$ plus and additional
$\so(n^{2.5}(\log(\norm{A}+\log\norm{b})))$ bit operations.  The
algorithm is probabilistic of the Las Vegas type.  That is, it assumes
the ability to generate random bits at unit cost, and always returns
the correct answer with controllably high probability.  When
$\phi(n)=\so(n)$, the implied cost of $\so(n^{2.5})$ bit operations
improves upon the $p$-adic lifting method of \cite{Dixon:1982:Pad}
which requires $\so(n^3)$ bit operations for sparse or dense matrices.
This theoretical efficiency was reflected in practice in
\cite{EGGSV06-1} at least for large matrices.

We present several other rather surprising applications of this
technique. Each incorporates the technique into an existing algorithm
in order to reduce the asymptotic complexity for the sparse matrix
problem to be solved. In particular, given a matrix $A\in\F^\nxn$ over an
arbitrary field $\F$, we are able to compute the complete inverse of
$A$ with $\so(n^{3-1/(\omega-1)})$ operations in $\F$ plus
$\so(n^{2-1/(\omega-1)})$ matrix-vector products by $A$.  Here
$\omega$ is such that we can multiply two $n\times n$ matrices with
$O(n^\omega)$ operations in $\F$.  Standard matrix multiplication
gives $\omega=3$, while the best known matrix multiplication of
Coppersmith \& Winograd \cite{CoWi90} has $\omega=2.376$.  If again we
can compute $v\mapsto Av$ with $\so(n)$ operations in $\F$, this
implies an algorithm to compute the inverse with
$\so(n^{3-1/(\omega-1)})$ operations in $\F$.  This is always less
than $O(n^\omega)$, and in particular equals $\so(n^{2.27})$
operations in $\F$ for the best known $\omega$ of \cite{CoWi90}.
Other relatively straightforward applications of these techniques
yield algorithms for the full nullspace and (certified) rank with this
same cost.  Finally, we sketch how these methods can be employed in
the algorithms of Kaltofen and Villard \cite{KalVil04} and \cite{Gie01}
to computing the determinant and Smith form of sparse matrices more
efficiently.

There has certainly been much important work done on finding exact
solutions to sparse rational systems prior to \cite{EGGSV06-1}.
Dixon's $p$-adic lifting algorithm \cite{Dixon:1982:Pad} performs
extremely well in practice for dense and sparse linear systems, and is
implemented efficiently in LinBox \cite{jgd:2002:icms}, Maple and
Magma (see \cite{EGGSV06-1} for a comparison).  Kaltofen \& Saunders
\cite{KalSau91} are the first to propose to use Krylov-type algorithms
for these problems.  Krylov-type methods are used to find Smith forms
of sparse matrices and to solve Diophantine systems in parallel in
\cite{Gie97,Gie01}, and this is further developed in
\cite{KalVil04,DumSauVil00}.  See the references in these papers for a
more complete history.  For sparse systems over a field, the seminal
work is that of Wiedemann \cite{Wiedemann:1986:SSLE} who shows how to
solve sparse $n\times n$ systems over a field with $O(n)$
matrix-vector products and $O(n^2)$ other operations.  This research
is further developed in \cite{KalSau91,cekstv:02:laa,Ka:95:mathcomp}
and many other works.  The bit complexity of similar operations for
various families of structure matrices is examined in \cite{EmiPan05}.


\section{Sparse block generators for \protect\linebreak the Krylov space}

For now we will consider an arbitrary invertible matrix $A\in\F^\nxn$
over a field $\F$, and $s$ an integer --- the blocking factor ---
which divides $n$ exactly.  Let $m=n/s$.  For a so-called
\emph{block projection} $u \in \F ^{n \times s}$ and $1\leq k \leq m$, we
denote by $\KK_k(A,u)$ the \emph{block Krylov matrix} $[u, Au, \ldots,
A ^{k-1}u] \in \F^{n \times ks}$.  Our goal is to show that
$\KK_m(A,u)\in\F^\nxn$ is non-singular for a particularly simple and
sparse $u$, assuming some properties of $A$.
 
Our factorization uses the special projection (which we will refer to
as an \emph{efficient block projection})
\begin{equation} 
  \label{eq:defu}
  u = \left[
    \begin{array}{cccc}
      I_s \\
      \vdots \\
      I_s
    \end{array}
  \right] \in \F ^{n \times s}
\end{equation}
which is comprised of $m$ copies of $I_s$ and thus has exactly
$n$ non-zero entries.  A similar projection has been suggested
in~\cite{EGGSV06-1} without proof of its reliability (i.e., that the
corresponding block Krylov matrix is non-singular).  We establish here
that it does yield a block Krylov matrix of full rank, and hence can be
used for an efficient inverse of a sparse $A$.

Let $\DD = \diag(\delta_1,\ldots,
\delta_1,\delta_2,\ldots,\delta_2,\ldots, \delta_m, \ldots,
\delta_m)$ be an $n \times n$ diagonal matrix whose entries consist of
$m$ distinct indeterminates $\delta_i$,
each $\delta_i$ occurring $s$ times.
\begin{theorem} 
  \label{thm:invK} 
  If the leading $ks \times ks$ minor of $A$ is non-zero 
  for $1\leq k\leq m$, then $\KK_m(\DD A \DD,u)
  \in \F ^{n \times n}$ is invertible.
\end{theorem}
\begin{proof}
  Let $\BB=\DD A \DD$.  
  For $1 \leq k \leq m$, 
  define $\BB_k$ as the specialization of $\BB$ 
  obtained by setting $\delta_{k+1},\delta_{k+2},\ldots,\delta_m$ to zero.
  Then $\BB_k$ is the matrix constructed by setting to zero the
  last $n-ks$ rows and columns of $\BB$.
  Similarly, for $1 \leq k \leq m$ we define $u_k \in \F^{n \times s}$
  to be the matrix constructed from
  $u$ by setting to zero the last $n-ks$ rows.
  In particular we have $\BB_m=\BB$ and $u_m=u$.

  We proceed by induction on $k$ and show that
  \begin{equation}
    \label{eq:recuK} 
    \rank \KK_k(\BB_k,u_k)= ks,
  \end{equation}
  for $ 1 \leq k \leq m$.  
  For the base case $k=1$ we have $\KK_1(\BB_1,u_1) = u_1$ and 
  thus $\rank \KK_1(\BB_1,u_1) = \rank u_1 = s$.

  Now, assume that \eqref{eq:recuK} holds for some $k$ with $1 \leq k
  < m$. By the definition of $\BB_k$ and $u_k$, 
  only the first $ks$ rows of $\BB_k$ and $u_k$ 
  will be involved in the left hand side of (\ref{eq:recuK}).
  Similarly, only the first $ks$ columns of $\BB_k$ will be involved.
  Since by assumption on $\BB$ the leading $ks \times ks$
  minor is non-zero, we have
  $\rank \BB_k \KK_k(\BB_k,u_k)
  = ks$, which 
  is equivalent to
  $\rank \KK_k(\BB_k,\BB_k u_k)=ks.$
  By the fact that the first $ks$ 
  rows of $u_{k+1}-u_k$ are zero, we have $\BB_k (u_{k+1}-u_k)=0$,  
  or equivalently $\BB_k u_{k+1} = \BB_ku_k$, and hence
  \begin{equation} 
    \label{eq:recuBK} 
    \rank \KK_k(\BB_k,\BB_k u_{k+1}) = ks.
  \end{equation}
  The matrix in (\ref{eq:recuBK}) can be written as
  \[ 
  \KK_k(\BB_k,\BB_k u_{k+1}) = \left [
     \begin{array}{c} M_k \\ 0 \end{array} \right ] \in \F^{n \times ks},
  \]
  where $M_k  \in \F^{ks \times ks}$ is nonsingular.
  Introducing the block $u_{k+1}$ we obtain the following
matrix:
  \begin{equation} 
    \label{eq:recuBKa} 
    \left [ u_{k+1}, \KK_k(\BB_k,\BB_k u_{k+1})\right ] =
   \left[ \begin{array}{cc}
      \ast & M_k  \\
      I_s & 0 \\
      0 & 0 
    \end{array}
  \right].
  \end{equation}
  whose rank is $(k+1)s$.  Noticing that 
  \[
  \biggl[u_{k+1}, \KK_k(\BB_k,\BB_k u_{k+1})\biggr] =
  \biggl[\KK_{k+1} \left(\BB_k, u_{k+1}\right)\biggr],
  \]
  we are led to
  \[
  \rank \KK_{k+1}(\BB_k, u_{k+1})
   = (k+1)s. 
  \]
  Finally, using the fact that 
  $\BB_k$ is the specialization of $\BB_{k+1}$ obtained by setting $\delta_{k+1}$
to zero, we obtain
  \[
    \rank \KK_{k+1}(\BB_{k+1}, u_{k+1})
     = (k+1)s,
    \]
  which is~\eqref{eq:recuK} for $k+1$ and thus establishes the theorem
  by induction.
\end{proof}

If the leading $ks \times ks$ minor of~$A$ is nonzero then the leading
$ks \times ks$ minor of~$A^T$ is nonzero as well, for any
integer~$k$.

\begin{corollary}
\label{cor:invK}
If the leading $ks \times ks$ minor of~$A$ is nonzero for $1 \le k \le
m$ and $\BB = \DD A \DD$, then $\KK_m(\BB^T, u) \in \F^{n \times n}$
is invertible. 
\end{corollary}

Suppose now that $A \in \F^{n \times n}$ is an arbitrary non-singular
matrix and the size of~$\F$ exceeds~$n(n+1)$. It follows by Theorem~2
of Kaltofen and Saunders~\cite{KalSau91} that there exists a
lower triangular Toeplitz matrix $L \in \F^{n \times n}$ and an upper
triangular Toeplitz matrix $U \in \F^{n \times n}$ such that each of
the leading minors of $\widehat{A} = UAL$ is nonzero. Let $\BB = \DD
\widehat{A} \DD$; then the products of the determinants of the
matrices $\KK_m(\BB, u)$ and $\KK_m(\BB^T, u)$ (mentioned in the above
theorem and corollary) is a polynomial with total degree less than
$2n(m-1) < n(n+1)$ (if $m \not= 1)$. In this case it follows that there is
also a non-singular diagonal matrix $D \in \F^{n \times n}$ such that
$\KK_m(B, u)$ and $\KK_m(B^T, u)$ are non-singular, for
\[
B = D \widehat{A} D = DUALD.
\]

Now let $R = LD^2 U \in \F^{n \times n}$, $\widehat{u} \in
\F^{s \times n}$ and $\widehat{v} \in \F^{n \times s}$ such that
\[
\widehat{u}^T = (L^T)^{-1} D^{-1} u
\quad \text{and} \quad
\widehat{v} = LDu.
\]
Then
\[
\KK_m(RA, \widehat{v}) = LD \KK_m(B, u)
\]
and
\[
L^T D \KK_m((RA)^T, \widehat{u}^T) = \KK_m(B^T, u),
\]
so that $\KK_m(RA, \widehat{v})$ and $\KK_m((RA)^T, \widehat{u}^T)$
are each non-singular as well. Since $D$ is diagonal and $U$ and~$L$
are triangular Toeplitz matrices it is now easily established that
$(R, \widehat{u}, \widehat{v})$ is an ``efficient block projection''
for the given matrix~$A$, where this is as defined
in~\cite{EGGSV06-1}.

This proves Conjecture~2.1 of~\cite{EGGSV06-1} for the case that the
size of~$\F$ exceeds $n(n+1)$:

\begin{corollary}
  \label{con:espexists}
  For any non-singular $A\in\F^\nxn$ and $s\divs n$ (over a field of
  size greater than $n(n+1)$) there exists an efficient block
  projection $(R,u,v)\in\F^\nxn\times \F^{s\times n}\times \F^{n\times
    s}$.
\end{corollary}


\section{Factorization of the matrix inverse}

The existence of the efficient block projection established in the
previous section allows us to define a useful factorization of the
inverse of a matrix.  This was used to obtain faster heuristics for
solving sparse integer matrices in \cite{EGGSV06-1}.  The basis is the
following factorization of the matrix inverse.

Let $\BB=\DD A \DD$, where $\DD$
is an $n\times n$ diagonal matrix whose diagonal entries consist of
$m$ distinct indeterminates, each occuring $s$ times contiguously, as
previously.  Define $\KKur=\KK_m(\BB,u)$ with $u$ as in
\eqref{eq:defu} and $\KKul=\KK_m(\BB^T,u)^T$ (where $(r)$ and $(\ell)$
refer to projection on the right and left respectively).  For any $0
\leq k \leq m-1$ and any two indices $l$ and $r$ such than $l+r=k$ we
have $u^T \BB^l \cdot \BB^r u = u^T \BB^k u$.  Hence the matrix
$\HH_u= \KKul\cdot \BB\cdot \KKur$ is block-Hankel with blocks of
dimension $s \times s$:
\[
\begin{aligned}
  \HH_u & =\left[\begin{array}{cccc}
      u^T \BB u &   u^T \BB^2 u & \ldots &  u^T \BB^m u\\
      u^T \BB2 u &   u^T \BB^3 u & \kern-28pt\rddots  &  \vdots \\
      \vdots  & \kern36pt\rddots   &  &  u^T \BB^{2m-2} u \\
      u^T \BB^m u&   \ldots  &   u^T \BB^{2m-2} u&  u^T \BB^{2m-1} u
    \end{array}\right]\\
  & \in \F^{n \times n} 
\end{aligned}
\]

From $\HH_u= \KKul\cdot \BB\cdot \KKur = \KKul\cdot \DD A \DD \cdot
\KKur$, we have following corollary to Theorem~\ref{thm:invK}, which
(together with Corollary~\ref{cor:invK})
implies that $\KKur$ and $\KKul$, and thus $\HH_u$ are invertible.

\begin{corollary}
  If $A\in\F^\nxn$ is such that all leading $ks\times ks$ minors are
  non-singular, $\DD$ is a diagonal matrix of indeterminates and
  $\BB=\DD A \DD$, then $\BB^{-1}$ and $A^{-1}$ may be factored as
  \begin{equation} 
    \label{eq:fact} 
    \begin{aligned}
      \BB^{-1} = & \KKur \HH_u^{-1} \KKul, \\
      A^{-1} = & \DD \KKur \HH_u^{-1} \KKul \DD,
      \end{aligned}
  \end{equation}
  where $\KKul$ and $\KKur$ are block-Krylov as defined above, and
  $\HH_u\in\F^\nxn$ is block-Hankel (and invertible) with $s \times s$
  blocks, as above.
\end{corollary}
We note that for any specialization of the indeterminates in $\DD$
to field elements in $\F$ such that $\det\HH_u\neq 0$ we get a similar
formula to \eqref{eq:fact} completely over $\F$.  A similar
factorization in the non-blocked case is used in~\cite[(4.5)]{Ebe97}
for fast parallel matrix inversion.




\section{Black-box matrix inversion over a field}
 
Let $A\in\F^\nxn$ be invertible and such that for any $v \in
\F^{n\times 1}$ the matrix times vector product $Av$ or $A^Tv$ can be
computed in $\phi(n)$ operations in $\F$ (where $\phi(n) \geq n$).
Following Kaltofen, we call such matrix-vector and vector-matrix
products \emph{black-box} evaluations of $A$.  In this section we will
show how to compute $A^{-1}$ with $\so(n^{2-1/(\omega-1)})$ black box
evaluations and additional $\softO(n^{3-1/(\omega-1)})$ operations in
$\F$.  Note that when $\phi(n)=\so(n)$ --- a common characterization
of ``sparse'' matrices --- the exponent in $n$ of this cost is smaller
than $\omega$, and is $\so(n^{2.273})$ with the currently best-known
matrix multiplication.

We assume for the moment that all principal $ks\times ks$ minors of
$A$ are non-zero, $1\leq k\leq m$.

Let $\delta_1, \delta_2, \ldots, \delta_m$ be the indeterminates
which form the diagonal entries of $\DD$ and let $\BB = \DD A \DD$.
By Theorem~\ref{thm:invK} and Corollary~\ref{cor:invK}, the matrices
$\KK_m(\BB, u)$ and~$\KK_m(\BB^T, u)$ are each invertible.
If $m \ge 2$ then the product of the determinants of these matrices is
a non-zero polynomial $\Delta \in \F [\delta_1,\ldots,\delta_m]$ with
total degree less than $2n(m-1)-1$.

Suppose that $\F$ has at least $2n(m-1)$ elements. Then $\Delta$
cannot be zero at all points in $(\F \setminus \{0\})^n$. Let $d_1,
d_2, \dots, d_m$ be nonzero elements of~$\F$ such that $\Delta(d_1,
d_2, \dots, d_m) \not= 0$, let
$D=\diag(d_1,\ldots,d_1,\ldots,d_m,\ldots,d_m)$, and let $B=DAD$. Then
$\Kur=\KK_m(B,u)\in\F^\nxn$ and $\Kul=\KK_m( B^T,u)^T\in\F^\nxn$ are
each invertible since $\Delta(d_1, d_2, \dots, d_m) \not= 0$, $B$ is
invertible since~$A$ is and $d_1, d_2, \dots, d_m$ are all nonzero,
and thus $H_u=\Kul B \Kur\in\F^\nxn$ is invertible as well.  Correspondingly,
\eqref{eq:fact} suggests
\[
B^{-1} = \Kur H_u^{-1} \Kul, \quad \mbox{and}~~ A^{-1} = D \Kur
H_u^{-1} \Kul D
\]
for computing the matrix inverse.

\begin{description}
\item[1. Computation of $u^T$, $u^TB$, \ldots, $u^TB^{2m-1}$ and
  $\Kul$.]
  We can compute this sequence, and hence $\Kur$ with $m-1$
  applications of $B$ to vectors using $O(n\phi(n))$ operations in
  $\F$.

\item[2. Computation of $H_u$.]\ \\
  Due to the special form \eqref{eq:defu} of $u$, one may then compute
  $wu$ for any $w\in\F^{s\times n}$ with $O(sn)$ operations.  Hence we
  can now compute $u^TB^iu$ for $0\leq i\leq 2m-1$ with $O(n^2)$
  operations in $\F$.

\item[3. Computation of $H_u^{-1}$.]\ \\
  The off-diagonal inverse representation of $H_u^{-1}$ as in
  \eqref{eq:Hoffdiag} in the Appendix can be found with $\so(s^\omega
  m)$ operations by Proposition~\ref{prop:costinvH}.

\item[4. Computation of $H_u^{-1}\Kul$.]\ \\
  From Corollary~\ref{cor:prodinvH} in the Appendix, we can compute
  the product $H_u^{-1}M$ for \emph{any} matrix $M\in\F^\nxn$ with
  $\so(s^{\omega}m^2)$ operations.

\item[5. Computation of $\Kur\cdot (H_u^{-1}\Kul)$.]\ \\
  We can compute $\Kur M=[u, Bu, \ldots, B^{m-1}u]M$, for
  \emph{any} $M\in \F^{n \times n}$ by splitting $M$ into $m$ blocks of
  $s$ consecutive rows $M_i$, for $0 \leq i \leq m-1$:
  \begin{equation}
    \label{eq:KuM}
  \begin{split}
    K_u M = & \sum_{i=0}^{m-1} B^{i}(uM_i) \\
    = & uM_0 + B (uM_1 + B(uM_2 + \cdots \\
      & \hspace*{15pt}\cdots + B(uM_{m-2} + BuM_{m-1})\cdots).
  \end{split}
  \end{equation}
  Because of the special form \eqref{eq:defu} of $u$, each product
  $uM_i \in \F^{n \times n}$ requires $O(n^2)$ operations, and hence
  all such products involved in \eqref{eq:KuM} can be computed in
  $O(mn^2)$ operations.  Since applying $B$ to an $n \times n$ matrix
  costs $n \phi(n)$ operations, $\Kur M$ is computed in $O(mn
  \phi(n)+mn^2)$ operations using the iterative form of~(\ref{eq:KuM})
\end{description}

In total, the above process requires $O(mn)$ applications of $A$ to a vector
(the same as for $B$), and
$O(s^{\omega} m^2+mn^2)$ additional operations. 
If $\phi(n) = \so(n)$ the overall number of field operations
is minimized 
with the blocking factor $s=n^{1/(\omega-1)}$.



\begin{theorem}
  \label{thm:spinv}
  Let $A \in \F^{n \times n}$, where $n=ms$, be such that all leading
  $ks\times ks$ minors are non-singular for $1\leq k\leq m$.  Let $B=DAD$ for
  $D = diag(d_1, \dots, d_1, \dots, d_m, \dots, d_m)$ such
  that $d_1, d_2, \dots, d_m$ are nonzero and each of the matrices
  $\KK_m(DAD, u)$ and $\KK_m((DAD)^T, u)$ is invertible.
  Then the inverse matrix $A^{-1}$ can be
  computed using $O(n^{2-1/(\omega -1)})$ applications of $A$ to
  vectors and an additional $\so(n^{3-1/(\omega-1)})$ operations
  in~$\F$.
\end{theorem}

The above discussion makes a number of assumptions.

First, it assumes that the blocking factor $s$ exactly divides
$n$. This is easily accommodated by simply extending $n$ to the
nearest multiple of $s$, placing $A$ in the top left corner of the
augmented matrix, and adding diagonal ones in the bottom right corner.

Theorem \ref{thm:spinv} also makes the assumptions that all the
leading $ks\times ks$ minors of $A$ are non-singular and that the
determinants of $\KK_m(DAD, u)$ and $\KK_m((DAD)^T, u)$ are each nonzero.
While we know of no way to ensure this deterministically in the times
given, standard techniques can be used to obtain these properties
probabilistically if~$\F$ is sufficiently large.

Suppose, in particular, that $n \ge 16$ and that $\#F> 2(m+1)n\lceil
\log_2 n \rceil$. Fix a set $\SS$ of at least $2(m+1)n \lceil \log_2 n
\rceil$ nonzero elements of~$\F$.  We can ensure that the leading $ks
\times ks$ minors of $A$ are non-zero by pre-multiplying by a
butterfly network preconditioner $U$ with parameters chosen uniformly
and randomly from $\SS$.  If $U$ is constructed using the generic
exchange matrix of~\cite[\S6.2]{cekstv:02:laa}, then it will use at
most $n \lceil
\log_2 n \rceil /2$ random elements from $S$, and
from~\cite[Theorem~6.3]{cekstv:02:laa} it follows that all leading $ks
\times ks$ minors of $\Atil = UA$ will be non--zero simultaneously with
probability at least $3/4$.  This probability of success can be made
arbitrarily close to $1$ with a choice from a larger $\SS$. We note
that $A^{-1}=\Atil^{-1}U$.  Thus, once we have computed $\Atil^{-1}$
we can compute $A^{-1}$ with an additional $\so(n^2)$ operations in
$\F$, using the fact that multiplication of an arbitrary $n \times n$
matrix by an $n\times n$ butterfly preconditioner can be done with
$\so(n^2)$ operations.

Once again let $\Delta$ be the products of the determinants of the
matrices $\KK_m(\DD A \DD, u)$ and $\KK_m((\DD A \DD)^T, u)$, so that
$\Delta$ is nonzero with total degree less than~$2n(m-1)$.
If we choose randomly selected values from $\SS$ for
$\delta_1,\ldots,\delta_m$, since $\#S \ge 2(m+1)n\lceil \log_2
n\rceil>4\deg \Delta$, the probability that $\Delta$ is zero at this
point is at most $1/4$ by the Schwartz-Zippel Lemma~\cite{Sch80,Zip79}.

In summary, for randomly selected butterfly preconditioner $B$, and
independently and randomly chosen values $d_1, d_2, \dots, d_m$
the probability that $\Atil=UA$ has non-singular leading $ks\times
ks$ minors for $1\leq k\leq m$ \emph{and} $\Delta(d_1, d_2, \dots,
d_m)$ is non-zero is at least $9/16>1/2$ when random choices are made
uniformly and independently from a finite subset~$\SS$ of $\F
\setminus \{0\}$ with size at least $2(m+1)n \lceil \log_2 n \rceil$.
%
%

When $\#\F \le 2(m+1)n\lceil \log_2 n \rceil$ we can easily construct
a field extension $\E$ of $\F$ which has size greater than
$2(m+1)n\lceil \log_2 n\rceil$ and perform the computation in that
extension.  Since this extension will have degree $O(\log_{\#F} n)$
over $\F$, it will add only a logarithmic factor to the final cost.
While we certainly do not claim that this is not of practical concern,
it does not affect the asymptotic complexity.

Finally, we note that it is easily checked whether the matrix returned
by this algorithm is the inverse of the input by using
n~multiplications by~$A$ by the columns of the output matrix and
corresponding each to the corresponding column of the identity
matrix. This results in a Las Vegas algorithm for computation of the
inverse of a black-box matrix with the cost as given above.

\begin{theorem}
  \label{thm:fastinv}
  Let $A\in\F^\nxn$ be non-singular.  Then the inverse matrix $A^{-1}$
  can be computed by a Las Vegas algorithm whose expected cost is
   $\so(n^{2-1/(\omega-1)})$
  applications of $A$ to a vector and $\so(n^{3-1/(\omega-1)})$ 
  additional operations in~$\F$.
\end{theorem}

\begin{table}[h]
  \label{tb:expon}
  \begin{center}
    \begin{tabular}{|c|c|c|c|}\hline
      $\omega$  & Black-box & Blocking & {\em Inversion}  \\
      & applications  & factor $s$ & {\em cost} \\\hline
      \vbox to 12pt{}
      \hspace*{2.5pt}3 \hspace*{13pt} (Standard) & 1.5 & 1/2 & $\so(n^{2.5})$ \\[3pt]
      2.807 (Strassen) & 1.446   & 0.553 & $\so(n^{2.446})$ \\[3pt]
      2.3755 (Cop/Win) & 1.273 & 0.728 & $\so(n^{2.273})$ \\[2pt]\hline 
    \end{tabular}
    \caption{Exponents of matrix inversion with a matrix $\times$
      vector cost $\phi(n)=\so({n})$.}
  \end{center}
\end{table}

\begin{remark}
  The structure~(\ref{eq:defu}) of the projection $u$ plays a central
  role in computing the product of the block Krylov matrix by a $n
  \times n$ matrix.  For a general projection $u\in \F^{n\times s}$,
  how to do better than a general matrix multiplication, i.e., how to
  take advantage of the Krylov structure for computing $K_uM$, appears
  to be unknown.
\end{remark}

\subsection*{Applying a Black-Box Matrix  Inverse to a Matrix}

The above method can also be used to compute $A^{-1}M$ for \emph{any}
matrix $M\in\F^\nxn$ with the same cost as in Theorem~\ref{thm:fastinv}. 
Consider the new step 1.5:

\begin{description}
\item[1.5. Computation of $\Kul\cdot M$.]\  \\
  Split $M$ into $m$ blocks of $s$ columns, so that
  $M=[M_0,\ldots,M_{m-1}]$ where $M_k\in\F^{n\times s}$.  Now consider
  computing $\Kul\cdot M_k$ for some $k\in\{0,\ldots, m-1\}$.  This
  can be accomplished by computing $B^iM_k$ for $0\leq i\leq m-1$ in
  sequence, and then multiplying on the left by $u^T$ to compute
  $u^TB^iM_k$ for each iterate.

  The cost for computing $\Kul M_k$ for a single~$k$ by the above process is
  $n-s$ multiplication of~$A$ to vectors and
  $O(ns)$ additional operations
  in $\F$.  The cost of doing this for all $k$ such that $0 \le k \le m-1$ is
  thus $m(n-s)<nm$ multiplications of~$A$ to vectors and $O(n^2)$
  additional operations. Since applying $A$ (and hence $B$) to an
  $n\times n$ matrix is assumed to cost $n\phi(n)$ operations in $\F$,
  $\Kul\cdot M$ is computed in $O(mn\phi(n)+mn^2)$ operations in $\F$
  by the process described here.
\end{description}

Note that this is the same as the cost of Step~5, so the overall cost
estimate is not affected.  Since Step 4 does not rely on any special
form for $\Kul$, we can replace it with a computation of
$H_u^{-1}\cdot(\Kul M)$ with the same cost.  We obtain the following:
\begin{corollary}
  \label{thm:fastappinv}
  Let $A\in\F^\nxn$ be non-singular and let $M\in\F^\nxn$ be any
  matrix.  We can compute $A^{-1}M$ using a Las Vegas algorithm
  whose expected cost is
  $\so(n^{2-1/(\omega-1)})$ multiplications of $A$ to vectors and
  $\so(n^{3-1/(\omega-1)})$ additional operations in~$\F$.
\end{corollary}

The estimates in Table \ref{tb:expon} apply to this
computation as well.


\section{Applications to Black-Box Matrices over a Field}

The algorithms of the previous section have applications in some
important computations with black-box matrices over an arbitrary field
$\F$.  In particular, we consider the problems of computing the
nullspace and rank of a black-box matrix.  Each of these algorithms
is probabilistic of the Las Vegas type.  That is, the output is
certified to be correct.

Kaltofen \& Saunders \cite{KalSau91} present algorithms for computing
the rank of a matrix and for randomly sampling the nullspace,
building upon work of Wiedemann \cite{Wiedemann:1986:SSLE}.  In
particular, they show that for random lower upper and lower triangular
Toeplitz matrices $U,L\in\F^\nxn$, and random diagonal $D$, that all
leading $k\times k$ minors of $\Atil=UALD$ are non-singular for $1\leq
k\leq r=\rank A$, and that if $f^\Atil\in\F[x]$ is the minimal
polynomial of $\Atil$, then it has degree $r+1$ if $A$ is singular
(and degree $n$ if $A$ is non-singular).  This is proven to be true
for any input $A\in\F^\nxn$, and for random choice of $U$, $L$ and
$D$, with high probability.  The cost of computing $f^\Atil$ (and
hence $\rank A$) is shown to be $O(n)$ applications of the black-box
for $A$ and $O(n^2)$ additional operations in $\F$.  However, no
certificate is provided that the rank is correct within this cost (and
we do not know of one or provide one here).  Kaltofen \& Saunders
\cite{KalSau91} also show how to generate a vector uniformly and
randomly from the nullspace of $A$ with this cost (and, of course, this
is certifiable with a single evaluation of the black box for $A$).  We
also note that the algorithms of Wiedemann and Kaltofen \& Saunders
require only a linear amount of extra space, which will not be the
case for our algorithms.

We first employ the random preconditioning of \cite{KalSau91}:
$\Atil=UALD$ as above.  \emph{We will thus assume in what follows that
  $A$ has all leading $i\times i$ minors non-singular for $1\leq i\leq
  r$.}  While an unlucky choice may make this statement false, this
case will be identified in our method.  Also assume that we have
computed the rank $r$ of $A$ with high probability.  Again, this will
be certified in what follows.

\begin{description}
\item[1. Inverting the leading minor.] \ \\
  Let $A_0$ be the leading $r\times r$ minor of $A$ and partition $A$
  as
  \[
  A=\begin{pmatrix}
    A_0 & A_1 \\
    A_2 & A_3
  \end{pmatrix}
  \]
  Using the algorithm of the previous section, compute $A_0^{-1}$.  If
  this fails, then the randomized conditioning or the rank estimate
  has failed and we either report this failure or try again with a
  different randomized pre-conditioning.  If we can compute
  $A_0^{-1}$, then the rank of $A$ is at least the estimated $r$.

\item[2. Applying the inverted leading minor.]\ \\
  Compute $A_0^{-1}A_1\in\F^{r\times (n-r)}$ using the algorithm of the
  previous section (this could in fact be merged into the first step).

\item[3. Confirming the nullspace.]\ \\
  Note that
  \[
  \begin{pmatrix}
    A_0 & A_1\\
    A_2 & A_3
  \end{pmatrix}
  \underbrace{\begin{pmatrix}
    A_0^{-1}A_1\\
    -I
  \end{pmatrix}}_{N}
  = 
  \begin{pmatrix}
    0\\
    A_2A_0^{-1}A_1-A_3
  \end{pmatrix}
  =0
  \]
  and the Schur complement $A_2A_0^{-1}A_1-A_3$ must be zero if the
  rank $r$ is correct.  This can be checked with $n-r$ evaluations of
  the black box for $A$.  We note that because of its structure, $N$
  has rank $n-r$.
\item[4. Output rank and nullspace basis.]\ \\
  If the Schur complement is zero, then output the rank $r$ and $N$,
  whose columns give a basis for the nullspace of $A$.  Otherwise,
  output fail (and possibly retry with a different randomized
  pre-conditioning).
\end{description}

\begin{theorem}
  \label{thm:rankandnullspace}
  Let $A\in\F^\nxn$ have rank $r$.  Then a basis for the nullspace of
  $A$ and rank $r$ of $A$ can be computed with an expected number of
  $\so(n^{2-1/(\omega-1)})$ applications of $A$ to a vector, plus an
  additional expected number of $\so(n^{3-1/(\omega-1)})$ operations
  in $\F$.  The algorithm is probabilistic of the Las Vegas type.
\end{theorem}


\section{Applications to sparse rational linear systems}

Given a non-singular $A\in\ZZ^\nxn$ and $b\in\ZZ^\nx1$, in
\cite{EGGSV06-1} we presented an algorithm and implementation to
compute $A^{-1}b$ with $\so(n^{1.5}(\log(\norm{A}+\norm{b})))$
matrix-vector products $v\mapsto A\bmod p$ for a machine-word sized
prime $p$ and any $v\in\ZZ_p^\nx1$ plus
$\so(n^{2.5}(\log(\norm{A}+\norm{b})))$ additional bit-operations.
Assuming that $A$ and $b$ had constant sized entries, and that a
matrix-vector product by $A\bmod p$ could be performed with $\softO(n)$
operations modulo $p$, the algorithm presented could solve a system
with $\so(n^{2.5})$ bit operations.  Unfortunately, this result was
conditional upon the unproven Conjecture 2.1 of \cite{EGGSV06-1}: the
existence of an efficient block projection.  This conjecture was
established in Corollary 2.3 of the current paper.  We can now
unconditionally state the following:

\begin{theorem}
  Given any invertible $A\in\ZZ^\nxn$ and $b\in\ZZ^\nx1$, we can
  compute $A^{-1}b$ using a Las Vegas algorithm. The expected number
  of matrix-vector products $v\mapsto Av\bmod p$ is in
  $\so(n^{1.5}(\log(\norm{A}+\norm{b})))$, and the expected number of
  additional bit-operations used by this algorithm is in
  $\so(n^{2.5}(\log(\norm{A}+\norm{b})))$.
\end{theorem}

\subsection*{Sparse integer determinant and Smith form}

The efficient block projection of Theorem \ref{thm:invK} can also be
employed relatively directly into the block baby-steps/giant-steps
methods of \cite{KalVil04} for computing the determinant of an integer
matrix.  This will yield improved algorithms for the determinant and
Smith form of a sparse integer matrix.  Unfortunately, the new
techniques do not obviously improve the asymptotic cost of their
algorithms in the case for which they were designed, namely, for
computations of the determinants of dense integer matrices.

We only sketch the method for computing the determinant here following
the algorithm in Section 4 of \cite{KalVil04}, and estimate its
complexity.  Throughout we assume that $A\in\ZZ^\nxn$ is non-singular
and assume that we can compute $v\mapsto Av$ with $\phi(n)$
\emph{integer} operations, where the bit-lengths of these integers are
bounded by $\so(\log(n+\norm{v}+\norm{A}))$.

\begin{description}
\item[Preconditioning and setup.] \ \\
  Precondition $A\leftarrow B=D_1UAD_2$, where $D_1$, $D_2$ are random
  diagonal matrices, and $U$ is a unimodular preconditioner from
  \cite[\S5]{Wiedemann:1986:SSLE}.  While we will not do the detailed
  analysis here, selecting coefficients for these randomly from a set
  $S_1$ of size $n^3$ is sufficient to ensure a high probability of
  success.  This preconditioning will ensure that all leading minors
  are non-singular and that the characteristic polynomial is
  squarefree with high probability (see \cite{cekstv:02:laa} Theorem
  4.3 for a proof of the latter condition).  From Theorem
  \ref{thm:invK}, we also see that $\KK_m(B,u)$ has full rank with
  high probability.

  Let $p$ be a prime which is larger than the a priori bound on the
  coefficients of the characteristic polynomial of $A$; this is
  easily determined to be $(n\log\norm{A})^{n+o(1)}$.  Fix a
  blocking factor $s$ to be optimized later, and assume $n=ms$.

\item[Choosing projections.] Let $u\in\ZZ^{n\times s}$ be an efficient
  block projection as in \eqref{eq:defu} and $v\in\ZZ^{n\times s}$ a
  random (dense) block projection with coefficients chosen from a set
  $S_2$ of size at least $2n^2$.

\item[Forming the sequence $\alpha_i=uA^iv\in\ZZ^{s\times s}$.]\ \\
  Compute this sequence for $i=0\ldots 2m$.  Computing all the $A^iv$
  takes $\so(n\phi(n)\cdot m\log\norm{A})$ bit operations. Computing
  all the $uA^iv$ takes $\so(n^2\cdot m\log\norm{A})$ bit operations.

\item[Computing the minimal matrix generator.] \ \\
  The minimal matrix generator $F(\lambda)$ modulo $p$ can be computed
  from the initial sequence segment $\alpha_0,\ldots,\alpha_{2m-1}$.
  See \cite[\S4]{KalVil04}. This can be accomplished with
  $\so(ms^\omega\cdot n\log\norm{A})$ bit operations.
 
\item[Extracting the determinant.] \ \\
  Following the algorithm in \cite[\S4]{KalVil04}, we first check if
  its degree is less than $n$ and if so, return ``failure''.
  Otherwise, we know $\det F^A(\lambda)=\det(\lambda I-A)$. Return
  $\det A=\det F(0) \bmod p$.

\end{description}

The correctness of the algorithm, and specifically the block
projections, follows from fact that $[u, Au,\ldots,A^{m-1}u]$ is of
full rank with high probability by Theorem \ref{thm:invK}.  Since the
projection $v$ is dense, the analysis of \cite[(2.6)]{KalVil04} is
applicable, and the minimal generating polynomial will have full
degree $m$ with high probability, and hence its determinant at
$\lambda=0$ will be the determinant of $A$.

The total cost is $\so((n\phi(n)m+n^2m+nms^\omega)\log\norm{A})$ bit
operations, which is minimized when $s=n^{1/\omega}$.  This yields an
algorithm for the determinant which requires
$\so((n^{2-1/\omega}\phi(n)+n^{3-1/\omega})\log\norm{A})$ bit
operations.  This is probably most interesting when $\omega=3$, where
it yields an algorithm for determinant which requires
$\so(n^{2.66}\log\norm{A})$ bit operations on a matrix with
pseudo-linear cost matrix-vector product.

We also note that a similar approach allows us to use the Monte Carlo
Smith form algorithm of \cite{Gie01}, which is computed by means of
computing the characteristic polynomial of random preconditionings of
a matrix.  This reduction is explored in \cite{KalVil04} in the dense
matrix setting. The upshot is that we obtain the Smith form with the
same order of complexity, to within a poly-logarithmic factor, as we
have obtained the determinant using the above techniques.  See
\cite[\S7.1]{KalVil04} and \cite{Gie01} for details.  We make no claim
that this is practical in its present form.


\appendix 
\section{Applying the inverse of a block-Hankel matrix}

In this appendix we address asymptotically fast techniques for
computing a representation of the inverse of a block Hankel matrix,
for applying this inverse to an arbitrary matrix.  The fundamental
technique we will employ is to use the off-diagonal inversion formula
of Beckermann \& Labahn \cite{bela2} and its fast variants
\cite{Giorgi:2003:issac}. An alternative to using the inversion
formula would be to use the generalization of the Levinson-Durbin
algorithm in \cite{Kaltofen:1995:ACB}.

For an integer $m$ that divides $n$ exactly with $s=n/m$, let

\begin{equation}
  \label{eq:defH}
  H=
  \left[\begin{array}{cccc}
      \alpha_0 &   \alpha_1 & \ldots &  \alpha_{m-1}\\
      \alpha_1  & \alpha_2 & \kern-10pt\rddots  &  \vdots \\
      \vdots  & \kern36pt\rddots   &  &  \alpha_{2m-2} \\
      \alpha_{m-1}&   \ldots  &    \alpha_{2m-2}& \alpha_{2m-1} 
    \end{array}\right] \in \F^{n \times n}
\end{equation}
be a non-singular block-Hankel matrix whose blocks are $s \times s$
matrices over $\F$, and let $\alpha_{2m}$ be arbitrary in $\F ^{s
  \times s}$.  We follow the lines of~\cite{Labahn:1990:BlockHankel}
for computing the inverse matrix $H^{-1}$.  Since $H$ is invertible,
the following four linear systems
(see~\cite[(3.8)-(3.11)]{Labahn:1990:BlockHankel})
\begin{equation} 
  \label{eq:HsolveR}
  \begin{aligned}
  & H \left[ q_{m-1}, \cdots, q_0 \right]^t
  = \left[ 0, \cdots, 0, I  \right] \in \F ^{n \times s}, \\
  & H \left[ v_m, \cdots, v_1  \right]^t 
  =
  - \left[ \alpha_{m},  \cdots \alpha_{2m-1} \alpha_{2m} \right] 
  \in \F^ {n \times s},
  \end{aligned}
\end{equation}
and 
\begin{equation} 
  \label{eq:HsolveL}
  \begin{array}{l}
    \left[ q^{*}_{m-1}~~\ldots ~~ q^{*}_{0}\right]
    H = \left[ 0~~\ldots ~~ 0 ~~ I\right] \in \F ^{s \times n},\\[0.24cm]
    \left[ v^{*}_{m}~~\ldots ~~ v^{*}_{1}\right]
    H = - \left[ \alpha_m~~\ldots ~~ \alpha_{2m-1} ~~ \alpha_{2m}\right] \in \F ^{s \times n}, 
  \end{array}
\end{equation}
have unique solutions given by the $q_k, q^{*}_k \in \F^{s\times s}$,
(for $0 \leq k \leq m-1$), and the $v_k, v^{*}_k \in \F^{s\times s}$
(for $1\leq k \leq m$).  Then we have
(see~\cite[Theorem\,3.1]{Labahn:1990:BlockHankel}):
\begin{equation}
  \label{eq:Hoffdiag}
  \begin{aligned}
    H^{-1} & =  
    \left[ 
      \begin{array}{cccc}
        v_{m-1} & \ldots & v_1 & \!\!I \\
        \vdots &  \kern-7pt\rddots & \kern-9pt\rddots &  \\
        v_1 & \kern-13pt\rddots &  &  \\
        I& &  &
      \end{array}
    \right]
    \left[ 
      \begin{array}{ccc}
        q^{*}_{m-1} & \!\!\ldots & \!\!q^{*}_0 \\
        &  \!\!\ddots &\!\!\vdots \\
        & &  \!\!q^{*}_{m-1}
      \end{array}
    \right] \\
    & \quad - 
    \left[ 
      \begin{array}{cccc}
        q_{m-2} & \ldots & q_0 & \!\!0 \\
        \vdots &  \kern-7pt\rddots & \kern-9pt\rddots &  \\
        q_0 & \kern-13pt\rddots &  &  \\
        0& &  &
      \end{array}
    \right]
    \left[ 
      \begin{array}{ccc}
        v^{*}_{m} & \ldots & v^{*}_1 \\
        &\ddots & \vdots\\
        & &   v^{*}_{m}
      \end{array}
    \right].
    \end{aligned}
\end{equation}
The linear systems~(\ref{eq:HsolveR}) and~(\ref{eq:HsolveL}) may also
be formulated in terms of matrix Pad\'e approximation problems. We
associate to $H$ the matrix polynomial $A=\sum _{i=0}^{2m} \alpha_i
x^i \in \F^{s \times s}[x]$.  The $s \times s$ matrix polynomials
$Q,P,Q^{*},P^{*}$ in $\F^{s \times s}[x]$ that satisfy
\begin{equation} 
  \label{eq:AsolveQ}
  \begin{aligned}
    A(&x)Q(x) \equiv P(x) + x^{2m-1} \bmod x^{2m},\\
    & \text{where $\deg Q \leq m-1$ and $\deg P \leq m-2$,}
    \\[0.24cm]
    Q^{*}&(x)A(x) \equiv P^{*}(x) + x^{2m-1} \bmod x^{2m},\\
    & \text{where $\deg Q^{*}\leq m-1$ and $\deg P^{*} \leq m-2$}
\end{aligned}
\end{equation}
are unique and provide the coefficients $Q = \sum _{i=0}^{m-1} q_i
x^i$ and $Q^{*} = \sum _{i=0}^{m-1} q^{*}_i x^i$ for constructing
$H^{-1}$ using~(\ref{eq:Hoffdiag})
(see~\cite[Theorem\,3.1]{Labahn:1990:BlockHankel}). The notation
``$\bmod~x^{i}$'' for $i \geq 0$ denotes that the terms of degree $i$ or
higher are forgotten.  The $s \times s$ matrix polynomials
$V,U,V^{*},U^{*}$ in $\F^{s \times s}[x]$ that satisfy
\begin{equation} 
  \label{eq:AsolveV}
  \begin{aligned}
    &A(x)V(x) \equiv U(x)  \bmod x^{2m+1},~V(0)=I, \\ 
    &\qquad \textbf{where $\deg V \leq m$ and $\deg U \leq m-1$,} \\[0.24cm]
    &V^{*}(x)A(x) \equiv U^{*}(x) \bmod x^{2m+1},~V^{*}(0)=I,\\
    &\qquad\textbf{where $\deg Q^{*} \leq m-1$ and  $\deg P^{*} \leq m-2$,}
    \end{aligned}
\end{equation}
are unique and provide the coefficients $V = 1+ \sum _{i=1}^{m} v_i
x^i$ and $Q^{*} = 1+\sum _{i=1}^{m} v^{*}_i x^i$ for
(\ref{eq:Hoffdiag}).

Using the matrix Pad\'e formulation, the matrices $Q$, $Q^{*}$, $V$,
and $V^{*}$ may be computed using the $\sigma$-basis algorithm
in~\cite{Beckermann:1994:MPA}, or its fast counterpart
in~\cite[\S2.2]{Giorgi:2003:issac} that uses fast matrix
multiplication. For solving~(\ref{eq:AsolveQ}), the $\sigma$-basis
algorithm with $\sigma = s (2m-1)$ solves
\begin{align*}
&[A~~-I] \left[ \begin{array}{c} \overline{Q} \\
    \overline{P} \end{array}\right] = R x^{2m-1} \bmod x^{2m},\\
&[\overline{Q}^{*}~~\overline{P}^{*}] \left[ \begin{array}{c} A \\
    -I \end{array}\right] = R^{*} x^{2m-1} \bmod x^{2m},
\end{align*}
with $\overline{Q},\overline{P},\overline{Q}^{*},\overline{P}^{*} \in
\F^{s \times s}[x]$ that satisfy the degree constraints $\deg
\overline{Q} \leq m-1, \deg \overline{Q}^{*} \leq m-1$, and $\deg
\overline{P}\leq m-2, \deg \overline{P}^{*} \leq m-2$.  The residue
matrices $R$ and $R^{*}$ in $\F^{s \times s}$ are non-singular, hence
$\overline{Q}R^{-1}$ and $(R^{*})^{-1}\overline{Q}^{*}$ are solutions
$\overline{Q}$ and $\overline{Q}^{*}$ for applying the inversion
formula~(\ref{eq:Hoffdiag}).  For~(\ref{eq:AsolveV}), the
$\sigma$-basis algorithm with $\sigma = s (2m+1)$ leads to
\begin{align*}
[A~~-I]& \left[ \begin{array}{c} \overline{V} \\
    \overline{U} \end{array}\right] = \bmod x^{2m+1},\\
[\overline{V}^{*}~~\overline{U}^{*}] &\left[ \begin{array}{c} A \\
    -I \end{array}\right] = \bmod x^{2m+1}
\end{align*}
with $\deg \overline{V} \leq m, \deg \overline{V}^{*} \leq m$, and
$\deg \overline{U}\leq m-1, \deg \overline{U}^{*} \leq m-1$.  The
constant terms $\overline{V}(0)$ and $\overline{V}^{*}(0)$ in $\F^{s
  \times s}$ are non-singular, hence
$V=\overline{V}(\overline{V}(0))^{-1}$ and $V^{*}=
(\overline{V}^{*}(0))^{-1}\overline{V}^{*}$ are solutions for
applying~(\ref{eq:Hoffdiag}).  Using Theorem~2.4
in~\cite{Giorgi:2003:issac} together with the above material we get
the following cost estimate.

\begin{proposition}
  \label{prop:costinvH}
  Computing the expression~(\ref{eq:Hoffdiag}) of the inverse of the
  block-Hankel matrix~(\ref{eq:defH}) reduces to multiplying matrix
  polynomials of degree~$O(m)$ in $\F^{s \times s}$, and can be done
  with $\so(s^{\omega}m)$ operations in $\F$.
\end{proposition}

Multiplying a block triangular Toeplitz or Hankel matrix in $\F^{n
  \times n}$ with blocks of size $s \times s$ by a matrix in
$\F^{n\times n}$ reduces to the product of two matrix polynomials of
degree~$O(m)$, and of dimensions $s \times s$ and $s\times n$.  Using
the fast algorithms in~\cite{CanKal91} or~\cite{BoSc05}, such a $s
\times s$ product can be done in $\so(s^{\omega}m)$ operations.  By
splitting the $s\times n$ matrix into $s \times s$ blocks, the $s
\times s$ by $s\times n$ product can thus be done in $\so(m \times
s^{\omega}m)= \so(s^{\omega}m^2)$ operations.

For $n=s^{\nu}$ let $\omega(1,1,\nu)$ be the exponent of the problem
of $s \times s$ by $s \times n$ matrix multiplication over $\F$. The
splitting considered just above of the $s\times n$ matrix into $s
\times s$ blocks, corresponds to taking $\omega(1,1,\nu) = \omega +
\nu -1 < \nu + 1.376$ ($\omega < 2.376$ due to~\cite{CoWi90}), with
the total cost $\so(s^{\omega(1,1,\nu)}m)= \so(s^{\omega}m^2)$.
Depending on $\sigma \geq 1$, a slightly smaller bound than $\nu +
1.376$ for $\omega(1,1,\nu)$ may be used due the matrix multiplication
techniques specifically designed for rectangular matrices
in~\cite{HuPa98}.  This is true as soon as $\nu \geq 1.171$, and gives
for example $\omega(1,1,\nu) < \nu + 1.334$ for $\nu=2$, i.e., for
$s=\sqrt{n}$.

\begin{corollary} 
  \label{cor:prodinvH}
  Let $H$ be the block-Hankel matrix of~(\ref{eq:defH}). If the
  representation~\eqref{eq:Hoffdiag} of $H^{-1}$ is given then
  computing $H^{-1}M$ for an arbitrary $M \in \F^{n \times n}$ reduces
  to four ${s \times s}$ by ${s \times n}$ products of polynomial
  matrices of degree $O(m)$.  This can be done with
  $\so(s^{\omega(1,1,\nu)}m)$ or $\so(s^{\omega}m^2)$ operations in
  $\F$ ($n=s^{\nu}=ms$).
\end{corollary}


\bibliography{eggsv}

\end{document}